# Rheotaxis of Active Droplets

Prateek Dwivedi[1], Atishay Shrivastava[1], Dipin Pillai[1] and Rahul Mangal[1,a)]

[1]Department of Chemical Engineering, Indian Institute of Technology Kanpur, Kanpur-208016, India

a) Author to whom correspondence should be addressed: mangalr@iitk.ac.in

**Abstract**

Rheotaxis is a well-known phenomenon among microbial organisms and artificial active colloids, wherein the swimmers respond to an imposed flow. We report the first experimental evidence of upstream rheotaxis by spherical active droplets. It is shown that the presence of a nearby wall and the resulting strong flow-gradient at the droplet level is at the root of this phenomenon. Experiments with optical cells of different heights reveal that rheotaxis is observed only for a finite range of shear rates, independent of the bulk flow-rate. We conjecture that the flow induced distortion of an otherwise isotropic distribution of filled/empty micelles around the droplet propels it against the flow. We also show that nematic droplets exhibit elastic stress-induced oscillations during their rheotactic flight. A promising potential of manipulating the rheotactic behavior to trap as well as shuttle droplets between target locations is demonstrated, paving way to potentially significant advancement in bio-medical applications.

**1. Introduction**

A unique feature of artificial swimmers(*1*, *2*) is their capability to harvest energy from their surroundings to perform mechanical motion analogous to biological microorganisms such as bacteria and sperm cells. Biological swimmers use their internal molecular motors to drive appendages/flagella attached to their surface, which results in their directed movement.(*3–9*) However, the movement of artificial swimmers does not involve moving parts. Instead, they exploit a local gradient generated by their non-uniform interactions with the surroundings. For example, Janus particles (JPs) use in-built asymmetric chemical patterning on their two halves to propel themselves through self-diffusophoresis(*10*, *11*), i.e., by inducing slip at the solid boundary due to a self-generated gradient in solute concentration. On the other hand, active droplets, which are isotropic droplets of one fluid dispersed in another immiscible fluid, generate spontaneous asymmetry in interfacial tension ($\gamma$) through different mechanisms such as change in surfactant activity via interfacial reactions(*12–15*), or adsorption-depletion of surfactants triggered by micellar solubilization.(*16–24*) The gradient in interfacial tension results in Marangoni stress at the interface that drives the fluid from low $\gamma$ towards high $\gamma$.(*16*, *19*) Since, there is no net external force acting on the system, to conserve the overall linear momentum, this interfacial flow propels the droplet towards the region of low interfacial tension. These artificial swimmers facilitate enhanced control over the motion of individual colloidal units, which otherwise exhibit random motion due to the thermal fluctuations induced Brownian motion. Therefore, in last few decades, these systems have attracted huge interest for their potential application as cargo carriers in microscopic domains,(*25*, *26*) sensing,(*27*) isolation of pathogens,(*28*) and environmental remediation.(*29*)

Since, the fundamental principle of motion in artificial self-propelled systems is the same as in micro-organisms, they successfully mimic several characteristics of both isolated and collective microbial motility. For example, artificial active matter systems have been shown to demonstrate short-time directed and long-time random behavior,(*10*) which is analogous to the run and tumble motion of bacteria and algae. Another parallel is their capability to sense and respond to the gradients present in their surroundings, a behavior known as "*taxis*". *Chemotaxis* refers to the response to the gradient in concentration of an external chemical stimulus. Bacteria are well known for their chemotactic motion in guiding them towards nutrient rich surroundings and fleeing away from poisons.(*30*) Similar behavior has been observed in sperm cells as well.(*31*) This mechanism is also critical in cell migration, which plays crucial role in several physiological and pathological processes, such as migration of neurons or lymphocytes. Using 5CB liquid crystal droplets in TTAB aqueous solution as a model system, Jin *et al.*(*32*) reported similar chemotactic behavior of active droplets, where the droplets moved preferentially toward surfactant-rich regions and away from surfactant-lean regions. In microscopic domains, microorganisms demonstrate a variety of non-equilibrium phenomena which originate from their interactions (hydrodynamic and chemical) with ambient fluid and nearby surfaces.(*33–41*) In particular, ambient fluid flow near a no-slip impenetrable boundary exposes the micro-organisms to sharp velocity gradients. Under these circumstances, they are known to demonstrate upstream *rheotaxis*.[7,42–45] Due to the asymmetric body shape of microorganisms, for example, the helical-shaped flagella in sperm cells and bacteria, the wall induced hydrodynamic interactions coupled with their active motility, results in their upstream swimming under external flow. Asymmetric active JPs have also been shown to demonstrate rheotactic behavior due to the interplay of their polarity and the external shear torque.(*46–48*) In all these studies on upstream rheotaxis, the asymmetric shape of the active swimmer was an essential ingredient in the underlying physics. For symmetric spherical active JPs, external flow has been reported to induce only cross-stream

migration.(*49*, *50*) To the best of our knowledge, the response of symmetric spherical active droplets in externally imposed flow has not been explored yet.

In this work, we employ a nematic liquid crystal 4-Cyano- 4'-pentylbiphenyl (5CB) as the oil droplet in aqueous solution of ionic tetradecyltrimethylammonium bromide (TTAB) surfactant, which propels due to micellar solubilization. Among several other possible combinations of oil droplets in aqueous ionic surfactant solutions(*12–16*, *18*, *19*, *51–53*) that exhibit self-propelled motion through this mechanism, choosing a nematic liquid crystal as the active oil droplet is the most popular as it aids in direct visualization of the flow-field inside the droplet. We therefore use it in our study as well. In this work, we provide the first experimental evidence of upstream rheotaxis exhibited by active spherical droplets. Through our careful experiments, we conjecture this behavior to be a consequence of the nearby wall induced shear significantly altering the otherwise isotropic solute concentration around the droplet. We also report on the oscillations in the upstream trajectories caused by the elastic distortions of the nematic droplet phase. Finally, we demonstrate a promising potential of the upstream rheotactic phenomenon to trap as well as shuttle cargo delivery at target locations, a potentially significant advancement for bio-medical applications.

## 2. Experimental Setup

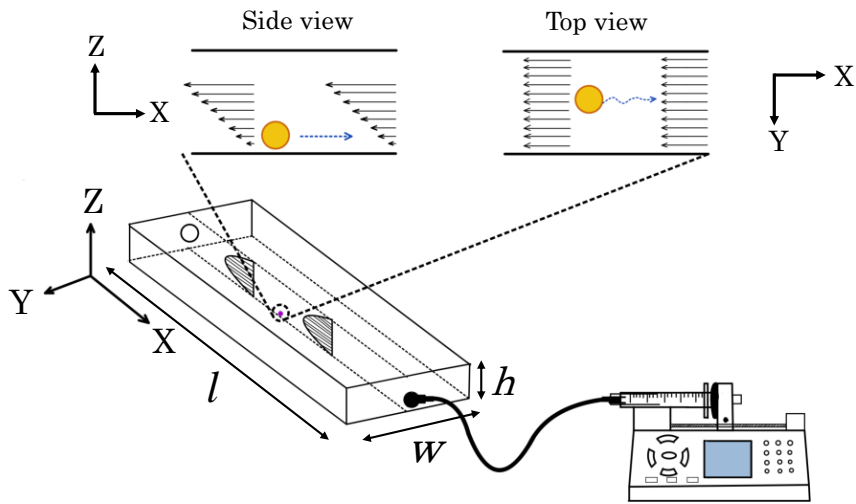

Figure 1. Schematic representation of experimental setup

To produce stable droplets of 5CB of size $R_d$~80 µm, a micro-injector (Femtojet 4i, Eppendorf) was used to inject appropriate amount of nematic liquid crystal 5CB into a petri-dish which was prefilled with 6 wt% aqueous solution of TTAB surfactant. The dilute emulsion of the 5CB droplets was then injected into a custom-made optical cell 10 (*w*) x 70 (*l*) mm$^2$ and vertical height *h* =3.2, 1.54 and 0.94 mm, prepared using glass slides which were thoroughly cleaned using the standard piranha solution. The cell was connected via tubing with a 10ml glass syringe prefilled with 6wt% TTAB aqueous surfactant solution. The syringe was mounted on a syringe pump (Chemyx 2000) using which the droplets in the cell were subjected to different volumetric flow rates (1, 2, 3, 5, 7, 8, 10, 12, 15, 20, 23, 30 and 50 ml hr$^{-1}$) along the length of the cell in the -X direction, as shown in the schematic figure 1. Since the 5CB droplets are denser ($\rho_{5CB}$~ 1.05 gm l$^{-1}$ and $\rho_{water}$~ 1.0 gm l$^{-1}$) than water, they readily sediment to the bottom wall of the cell. In absence of external flow, the droplets perform 2D motion while remaining confined to the X-Y plane parallel to the bottom wall. Exposure to shear flow occasionally lifts the droplet in the Z direction, i.e., normal to the flow direction and out of X-Y plane. The syringe pump generates a steady pressure driven parabolic flow, i.e., Poiseuille flow in the optical cell. However, since the droplets are mostly confined near the bottom wall and are much smaller in size compared to the cell height, the local flow profile around the droplets can be assumed to be a simple shear flow with gradient in the Z direction. An upright polarized optical microscope, Olympus BX53, was used to observe the active motion of the isolated droplets exposed to different flow rates. The motion of droplets was recorded at 20 fps using Olympus LC-30 camera with 1024 x 768 pixel$^2$ resolution mounted on the polarized optical microscope. To make the presence of side walls of cell irrelevant, droplets were focused within a narrow Y range (±500microns) along the centerline of the cell. Further, as the width of the cell in Y direction is much larger than the droplet size, flow can be considered uniform along the Y direction. Droplet tracking was performed with the Image-J software, using an image correlation-based approach, to obtain the particle trajectories in X-Y plane.

## 3. Results and Discussion
a) **Active motion in a quiescent medium**

We first benchmark our setup by conducting experiments to explore the droplet motion in the presence of 6wt.% TTAB aqueous solution under no external flow. As expected, the droplet performs random active motion in the 2D X-Y plane close to the bottom wall without any significant vertical (Z) drift. Figure 2(a) shows the 2-D mean squared displacement (MSD) of a representative active droplet trajectory ($\Delta t$ ~250 s), the inset shows the corresponding raw X-Y trajectory. It can be seen that at short times, the MSD scales as $\Delta t^2$, which is indicative of the droplet performing ballistic motion. However, at long time scales, MSD increases linearly with time, i.e., ~$\Delta t^1$. This is due to the persistent change in droplet's direction of locomotion, which eventually results in their random motion at longer time scales. However, the reason for persistent change in droplet direction is yet to be explored thoroughly. Our observations of droplet behavior in a quiescent medium are consistent with previous reports of Suga *et al.*(54) where droplets in the range of 50μm -100μm were shown to exhibit random motion at long time scales.

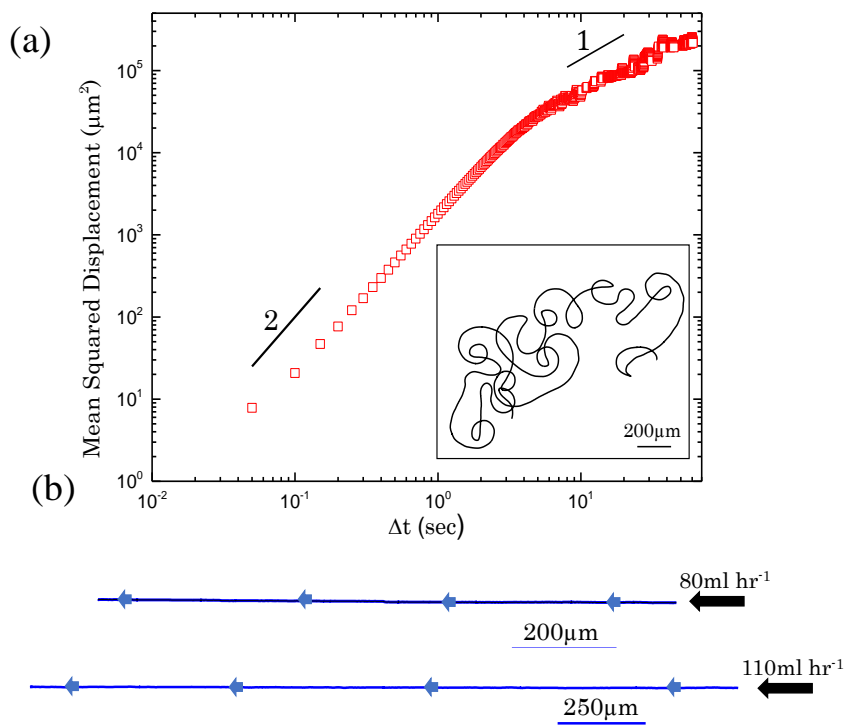

Figure 2. (a) Variation in Mean square displacement with time for corrosponding active 5CB droplet (radius ~80 μm) in an optical cell with *h*=3.2 mm . The inset shows the corresponding X-Y trajectory for 250 s (c) X-Y trajectory of passive 5CB droplets exposed to external shear flow in optical cells with *h*=3.2 mm

b) **Passive droplet in flow**

Next, as a control experiment, we analyze the motion of a passive 5CB droplet in external flow. To obtain passive 5CB droplets, we suspend them in 1 wt.% TTAB surfactant solution. At low flow rates (5-30 ml hr$^{-1}$), the droplet remains stationary, but at higher flow rates (80-110 ml hr$^{-1}$), the droplet moves along with the flow in -X direction, with insignificant drift in the orthogonal directions which can be seen from the droplet trajectories in figure 2(b).

c) **Active droplets in flow**

Now, we investigate the effect of external flow on the dynamics of an active 5CB droplet. The representative trajectories shown in figure 3 suggest that lower flow rates of 5 ml hr$^{-1}$ has negligible effect, and the droplet exhibits random motion without any directional bias. At intermediate flowrates of 10-20 ml hr$^{-1}$, the droplet unexpectedly moves against the flow in +X direction. This upstream migration is a well-known rheotactic behavior and has been well-studied for asymmetric micro-organisms(7, 55) and asymmetric active colloidal particles.(46, 48–50) Our study, to the best of our knowledge, is the first evidence of a spherically symmetric active system exhibiting upstream rheotaxis. By manipulating the rheotactic behavior through precise control of the external flow, we now demonstrate the possibility to trap as well as shuttle an active droplet between target locations. As shown in figure 4(a), we expose an active droplet to external flow-rates of 12 ml hr$^{-1}$ and 30 ml hr$^{-1}$ in a periodic manner. At 12 ml hr$^{-1}$, droplet performs upstream rheotaxis (red trajectory) and on subsequently changing the flow-rate to 30 ml hr$^{-1}$, it instantaneously reverses its direction and starts to drift with the flow (black trajectory) (c.f. figure 4(b)). During the second cycle, the droplet again resumes upstream rheotaxis at 12 ml hr$^{-1}$, followed by drifting with the flow at 30 ml hr$^{-1}$, albeit tracing a different trajectory this time. Thus, by periodically tuning the imposed flow-rate between low and high values, the droplet can be made to shuttle between desired target locations. Further, as droplet migration is found to reverse direction from positive to negative X direction with increase in flowrate, it is expected that there exists an intermediate flowrate that will result in no observable migration in X-direction. We show this indeed to be the case by using an intermediate flow-rate of 23 ml hr$^{-1}$, wherein the droplet oscillates purely in the Y direction, with negligible drift in the X direction (see figure 4(c)) These observations demonstrate that by careful manipulation of the imposed external flow, active droplets can be hydrodynamically trapped within targeted spatial domains to perform desired tasks.

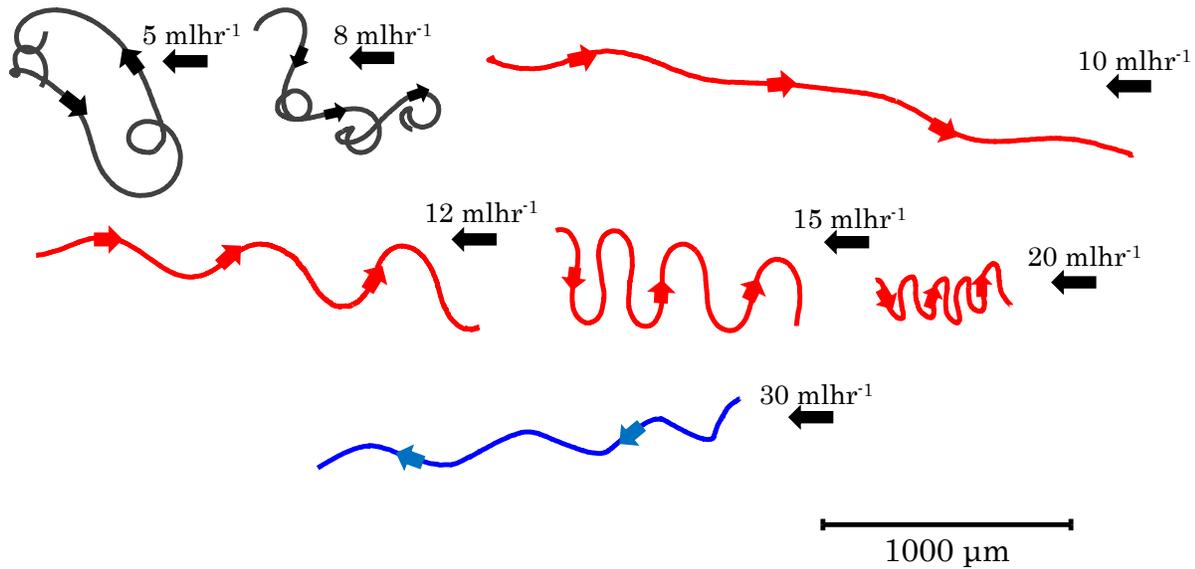

Figure 3. X-Y trajectories of active 5 CB droplets (Size ~80 μm) in an optical cell with $h$=3.2 mm exposed to varying external flow.

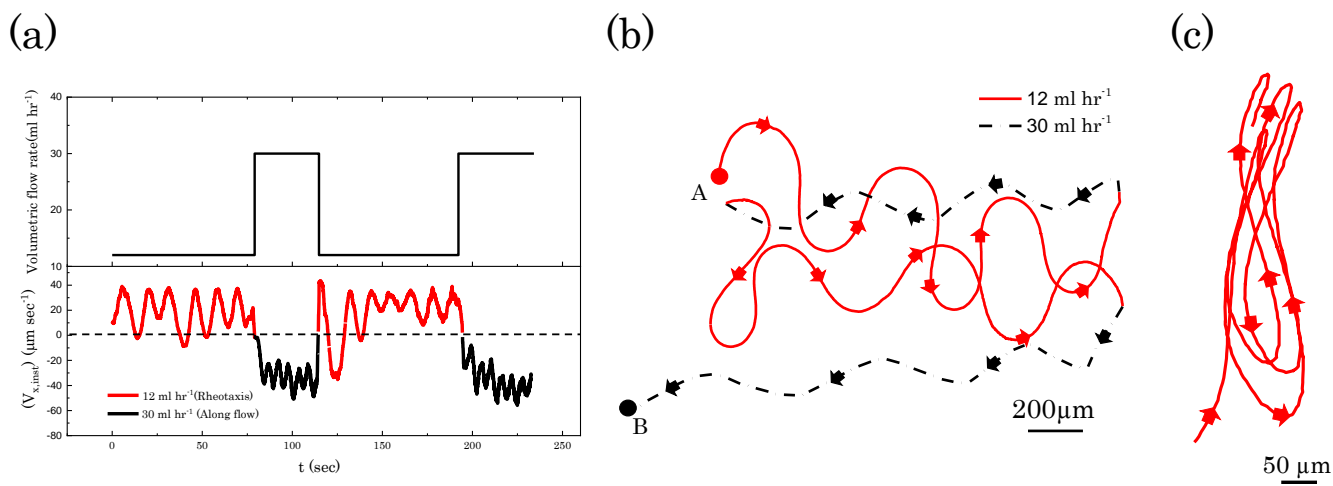

Figure 4. (a) Top panel: Variation of input flow-rate with time in a periodic manner, Bottom panel: Time variation of instantaneous X velocity of the active droplet (b) Corresponding X-Y trajectory starting at point A and ending at B. (c) X-Y trajectory of an oscillatory trajectory with insignificant X drift at external flow of 23 ml hr$^{-1}$.

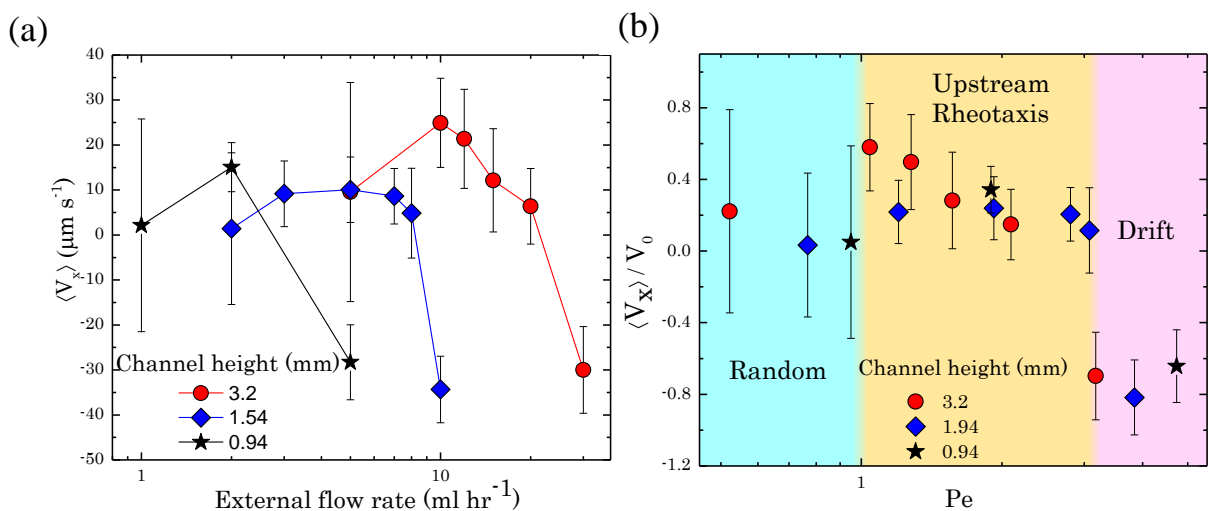

Figure 5. (a) Dependence of mean X velocity $\langle \mathbf{v}_x \rangle$ of the active 5CB droplets with respect to external flow-rate in optical cells with varying heights ($h$). (b) Variation in $\langle \mathbf{v}_x \rangle / V_o$ of the active 5CB dropelts with respect to non-dimensional $Pe$ in optical cells with varying heights.

To quantify the experimentally observed rheotactic motion, we calculate the average X velocity of the droplet, $\langle \mathbf{v}_x \rangle = \left\langle \dfrac{\mathbf{X}_{t+\Delta t} - \mathbf{X}_t}{\Delta t} \right\rangle$. Here, $\langle \ \rangle$ represents the time and ensemble (comprising of ~10-15 droplets) average. Figure 5(a) illustrates the variation in $\langle \mathbf{v}_x \rangle$ with respect to the flow rate in optical cells with varying vertical heights ($h$). As evident from the figure, for all optical cells, an increase in flow-rate results in a transition from active random motion ($\langle \mathbf{v}_x \rangle \sim 0$) → upstream rheotaxis ($\langle \mathbf{v}_x \rangle > 0$) → downstream drift with the flow ($\langle \mathbf{v}_x \rangle < 0$). Interestingly, in thinner cells, the transitions occur at lower flow-rates, highlighting the importance of the local shear-rate as compared to the absolute flow-rate. Based on the understanding of the mechanism behind upstream rheotaxis of mammalian sperm cells, *E.Coli* and artificial active colloids, we know that the interactions of the swimmer with the nearby wall plays a major role. Thus, to isolate the effect of the nearby wall, we repeated the flow experiments with heavy water (D$_2$O) added to the aqueous surfactant solution, so that the density of solution matches with the density of 5CB. In the density matched environment, active 5CB droplets no longer remain constrained to move along the 2D plane near the bottom wall, rather are free explore the Z direction as well. In these systems, in contrast to the previous case, on exposure to external flow, active 5CB droplets do not demonstrate upstream rheotactic behavior. At low flowrates of 0.3ml hr$^{-1}$ to 7ml hr$^{-1}$ droplets perform 3-D active random motion without any directional bias, and above 7ml hr$^{-1}$ droplets migrate with the flow while performing 3-D active motion.

In our experiments, since the droplet is located near the bottom wall and is much smaller than the cell height, it experiences the imposed flow as a simple shear flow with gradient in the Z direction. This suggests that the presence of the nearby wall and the consequent flow-gradient induced effects are essential for rheotaxis. In Figure 5(b), we plot the variation in $\langle \mathbf{v}_x \rangle / V_o$, with respect to a non-dimensional Peclet number, $Pe = \dot{\gamma}/D_{r,eff}$. The Peclet number, as defined here, is the ratio of the timescale associated with distortion of concentration field of micelles around the droplet to the timescale associated with the random unbiased motion of the droplet. Also, V$_o$ is the self-propelled speed of the active droplet in absence of flow and $\dot{\gamma} \sim |\mathbf{V}^*|/R$ is the local shear-rate. $\mathbf{V}^*$ is the theoretically obtained background flow velocity at the droplet center (see supporting discussion and figure S1). $D_{r,eff}$ is the rotational diffusion of the active droplet in the absence of imposed flow, calculated by fitting the temporal decay of the velocity autocorrelation $C(\Delta t) = \left\langle \dfrac{\mathbf{v}(t+\Delta t).\mathbf{v}(t)}{|\mathbf{v}(t+\Delta t)||\mathbf{v}(t)|} \right\rangle$ (supporting figure S2) with the following equation:

$$C(\Delta t) = 4D\delta(\Delta t) + \langle v \rangle^2 \cos(\omega \Delta t) \exp(-D_{r,eff}\Delta t) \tag{1}$$

Here, $\mathbf{v} = \dfrac{\mathbf{r}_{i+1} - \mathbf{r}_i}{t_{i+1} - t_i}$ is the instantaneous velocity vector, δ is the Dirac-delta function, and ω is the angular velocity which captures the curling behavior of the trajectory.

The plot suggests that for all cell heights, transitions in droplet behavior are well characterized by *Pe*, further reinforcing the importance of wall-induced shear. At low shear rates, (*Pe* <1), the flow is weak to distort the surfactant/filled micelle concentration around the droplet, and therefore the motion of the active droplet remains nearly unaltered. With increasing *Pe* (~1), the imposed shear rate successfully distorts droplets' neighborhood. The interplay of the droplets' activity and the modified flow causes the droplet to move against the flow. At very high *Pe* (>1), the imposed flow predominates their innate activity causing them to simply drift along the flow.

Based on the observations discussed above, we propose a possible mechanism underpinning the rheotactic behavior using the schematic illustration in figure 6. The self-propelling activity of the droplet is a consequence of micellar solubilization, wherein the local concentration of free surfactant molecules and filled micelles govern its motion.(*17*, *24*) Under quiescent conditions (cf. figure 6(a)), the droplet encounters free surfactants more or less at an equal rate everywhere along the droplet surface. Consequently, the local concentration (fresh surfactants and filled micelles) fluctuations are isotropic, resulting in random motion of the droplets. The presence of external flow breaks the spherical symmetry and an anisotropy in the supply of fresh surfactant molecules (along X direction in this case) is generated (cf. figure 6(b)). Further, the flow in X direction also sweeps away the filled micelles around the droplet interface. However, it is imperative to note that the droplet is adjacent to the bottom wall due its higher density. Therefore, the ambient velocity field has a strong gradient in the vertical Z direction with vanishing velocity near the bottom wall. Therefore, the supply of the fresh surfactants to the droplet interface will vary at different vertical locations. Likewise, the sweeping away of filled micelles will also vary at different vertical locations. Consequently, the downstream part of the droplet close to the bottom wall accumulates filled micelles with time due to the vanishing velocity there. As the filled micelles have to solely rely on diffusion to be swept away near the wall, the trail of filled micelles persists for a long time, promoting negative autochemotaxis and steering the droplet away from filled micelles. Thus, the asymmetric sweeping away of filled micelles creates an anisotropy across the droplet, which pushes the droplet against the flow.

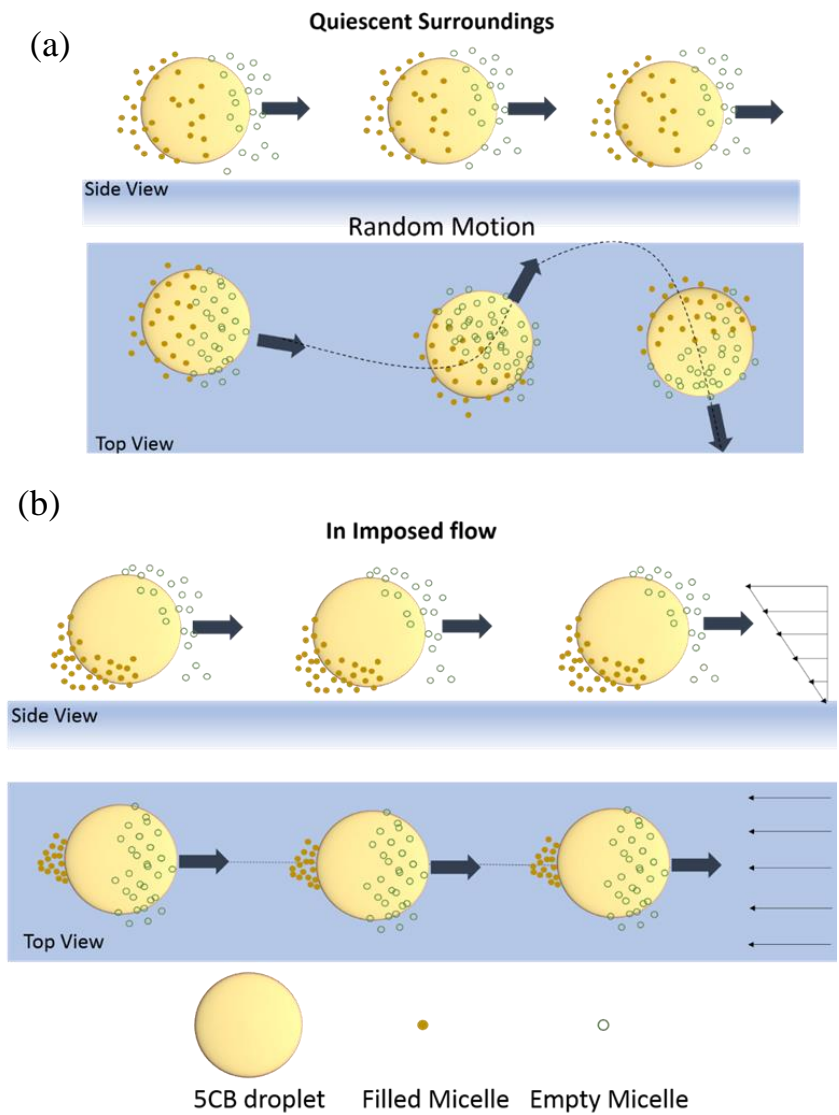

Figure 6. Proposed mechasnim of rheotaxis of active droplet near a planar wall. (a) Top and side view of the active droplet illustrating it's motion in absence of external flow. The concentration fluctuations of solutes around the droplet remains isotropic causing the droplet to change direction randomly. (b) Top and side view of the active droplet illustrating it's motion in presence of external flow.

Now we proceed to elucidate the oscillations observed in droplet trajectory during its rheotactic flight. At 10ml hr$^{-1}$, when the droplet just begins to perform upstream rheotaxis, the oscillations are weak, which then intensify with increase in flow-rate. With further increase in flow-rate, the oscillations become weak again. For instance, at 30 ml hr$^{-1}$, wherein the droplet advects downstream with the flow, the oscillations are weak. On subsequent increase of flow-rate to 50 ml hr$^{-1}$, the droplet migrates away from the bottom wall and drifts downstream with no oscillations, see figure 7. This non-oscillatory downstream drift at high flowrates, in fact, resembles that observed in control experiments with passive droplets at high flowrates. We speculate that at very high flowrates once the droplet migrates away from the bottom wall, the filled micelles are washed away more readily from the neighborhood of the droplet interface, resulting in negligible concentration gradient of filled micelles across the droplet. This causes the droplet to exhibit negligible activity in comparison to the imposed flow. A detailed analysis of the cross-stream migration of active 5CB droplets is beyond the scope of the current work and is reserved for future study. It should, however, be noted that cross-stream migration of surfactant-laden droplets under flow has been predicted in several earlier theoretical studies.(*56, 57*)

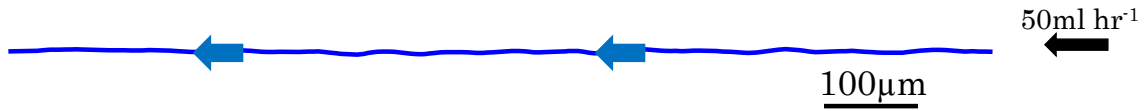

Figure 7. X-Y trajectory of the active droplet under external flow of 50 ml hr$^{-1}$ for duration of 15 s. (*Channel height h=3.2mm*)

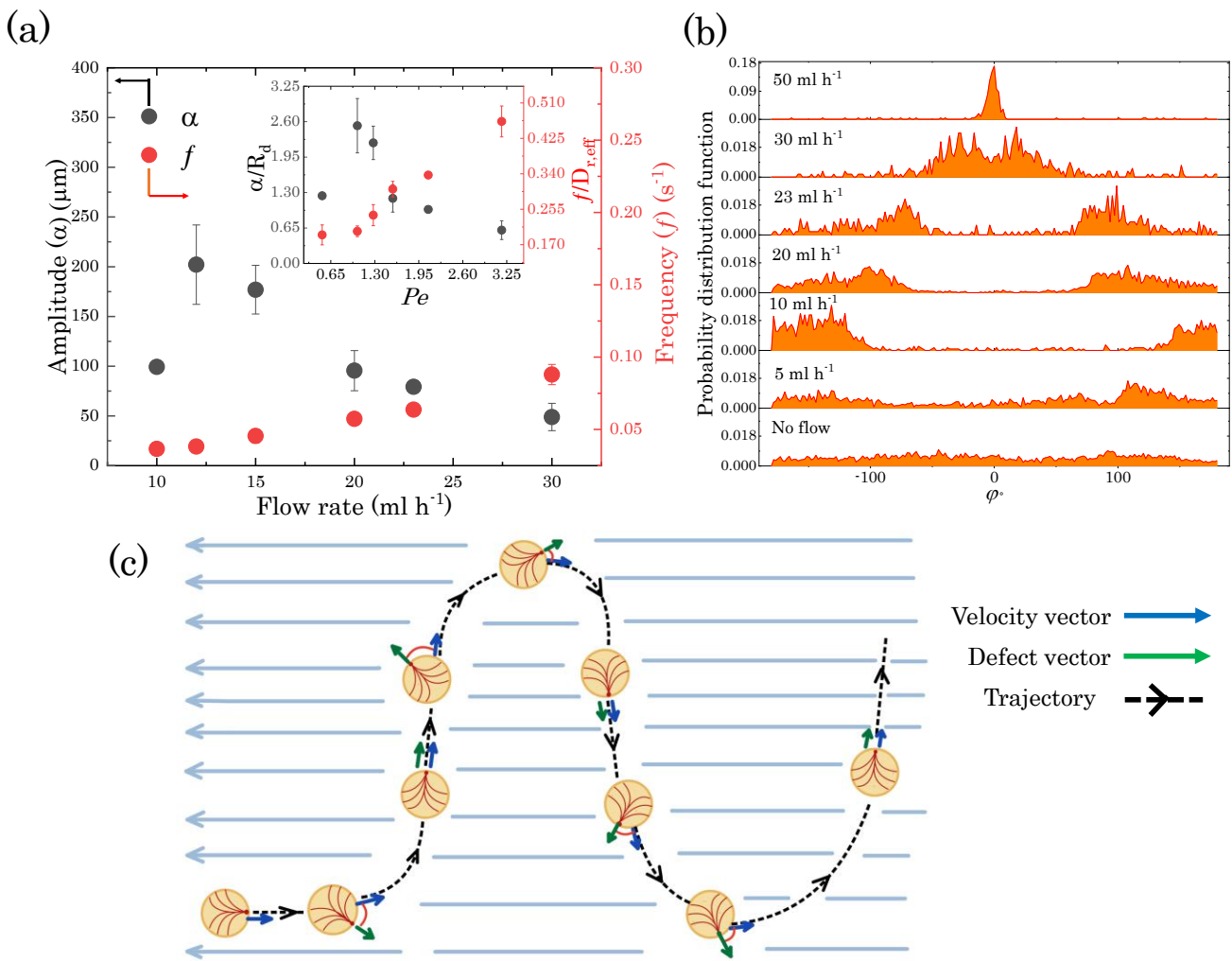

Figure 8. (a) Variation in the amplitude ($\alpha$) and oscillation frequency ($f$) with flow rate (*Channel height h=3.2mm*) for droplet trajectories. Inset shows the non-dimensionalized amplitude ($\alpha/R_d$) and oscillation frequency ($f/D_{r,eff}$) with $Pe$. (b) Probablity distribution of the velocity vector(*Channel height h=3.2mm*) ($\varphi$) (c) Proposed mechanism of oscillations during the rheotaxis motion of the active droplets.

Figure 8(a) depicts the oscillation amplitude $\alpha$ and frequency $f$ of the rheotactic trajectory for different flow-rates, obtained using Fourier transform of the droplet trajectory (y(x)). As evident from the figure, the maximum value of α (~200 μm) is insignificant compared to the lateral width (along Y) of the optical cell (~10000 μm), confirming that the oscillations are not due to the side wall confinement. The figure also shows that with increase in flow-rate, α initially increases followed by a decrease, while the $f$ increases monotonically. The inset of figure 8(a) depicts the non-dimensional amplitude ($\alpha/R_d$) and frequency ($f/D_{r,eff}$) as a function of Peclet number, $Pe$. The figure indicates that with increase in the Peclet number, the amplitude of oscillations become weak compared to the droplet size. It also highlights that the time-scale associated with the change in direction of the observed droplet trajectory, compared to their inherent rotational time-scale in absence of flow ($D_{r,eff}$), increases with the imposed flow strength. This facilitates the droplet to demonstrate biased oscillatory motion while performing rheotaxis in external flow, as opposed to random motion in a quiescent medium. Figure 8(b) illustrates the probability distribution P($\varphi$) of the orientation of the droplet velocity vector ($\varphi$) for different external flowrates. Under quiescent conditions, P($\varphi$) shows a nearly uniform distribution, characteristic of random motion. With increase in flow-rate to 20mlhr$^{-1}$, P($\varphi$) becomes a bimodal distribution with symmetric peaks at around ±90°, indicating that the motion is primarily directed in positive and negative Y-direction. On increasing the flow rate to 30 mlhr$^{-1}$, P($\varphi$) shows symmetric peaks at ~ ±25°, which is consistent with the downstream oscillatory drift of the active droplet. Here, the droplet activity is subdued by the external flow, which compels the droplet to migrate downstream. However, the presence of oscillations during the downstream migration is indicative of residual activity in the droplet. Imposed external flow forces the droplets to migrate downstream in -X direction with no Y drift. The residual activity results in small yet noticeable drift in Y direction. This perpendicular to flow drift generates an angle between droplets' velocity and external flow velocity which leads to oscillations, due to the dislocation of the nematic defect as discussed later. Due to the strong background flow, the oscillations are attenuated resulting in lower amplitude and higher frequency. A further increase in flowrate to 50 mlhr$^{-1}$ results in the droplet oscillation being totally suppressed in its downstream drift, as evident from the unimodal peak for P($\varphi$) at 0°.

In 2016, *Kruger et al.*(58) first reported a curling (2D)/helical (3D) behavior of active 5CB liquid crystal (LC) droplets in a quiescent aqueous TTAB solution. They showed that for a static LC droplet, the presence of adsorbed surfactants on the interface results in a homeotropic anchoring of the LC molecules near the surface, resulting in the nematic director field near the interface to point radially outward. As a result, a radial hedgehog point defect is observed at the center of the droplet. When the droplet self-propels, the viscous drag convects the defect to the front apex of the droplet due to internal flow within the droplet. The apex, however, is an unstable equilibrium for the defect, since any small fluctuation to either side results in the defect being convected further in that direction by the internal flow. Thus, the defect is deflected along the interface from the apex of the droplet towards its equator with equal probability in either direction. This deflection is however opposed by the elastic forces of the nematic LC and an equilibrium deflection angle ($\theta_s$) between the nematic director (**n**) and the velocity field is attained. The defect significantly lowers the nematic order in its neighborhood, as compared to other regions on the interface, where a radially normal director field due to homeotropic anchoring of the LC molecules persists. Therefore, while the effective viscosity for the rest of the interface is given by the Miesowicz second (also the largest) viscosity coefficient(*59*), the effective viscosity near the defect is given by a lower isotropic viscosity coefficient. Thus, an asymmetry in Marangoni flow is generated with stronger flow near the defect, and to conserve the overall angular momentum (on a torque free system), a rotational component opposite to the direction of deflection of the defect is added to the motion, which causes the droplet to change direction in a curling manner. We believe that a similar underlying mechanism is responsible for the oscillatory behavior of active droplets during their rheotactic flight.

In our experiments, for low external flow-rates, the viscous stresses originate solely due to droplets' self-propelled speed ($V_o$), which, for 80 μm droplets remain insignificant to invoke any elastic forces (~$1/R^2$), consistent with existing reports.(*54*) However, in experiments with flow, the interfacial viscous stresses are governed by the relative speed, $|V_{rel.}|=|V_o+V^*|$. For droplets migrating against the flow, $|V_{rel.}|>V_o$, and hence the viscous stresses are significantly higher, generating enough deflection in defect location that elastic forces become important causing the trajectories to curl. Unlike the quiescent medium of *Kruger et al.*(*58*), in our experiments, the interplay of the orientation of background flow with that of the droplet's self-propelled motion results in a periodic deflection of the point defect. As shown in the schematic in figure 8(c), during rheotaxis when the deflection of the defect is in the clock-wise direction, the trajectory will curl in the anti-clockwise direction. At the end of the first curl, the orientation of droplet's active motion with respect to the background flow is now reversed, such that defect now deflects in the opposite direction, forcing the droplet to switch the curling direction. This continued behavior leads to the curling induced oscillations. With increasing imposed flowrate, $|V_{rel.}|$ increases monotonically, and as a result $\theta_s$ also increases, causing enhanced oscillations. For high flowrates, when the droplet stops upstream migration, $|V_{rel.}|$ drops again, resulting in weaker oscillations in the downstream drift.

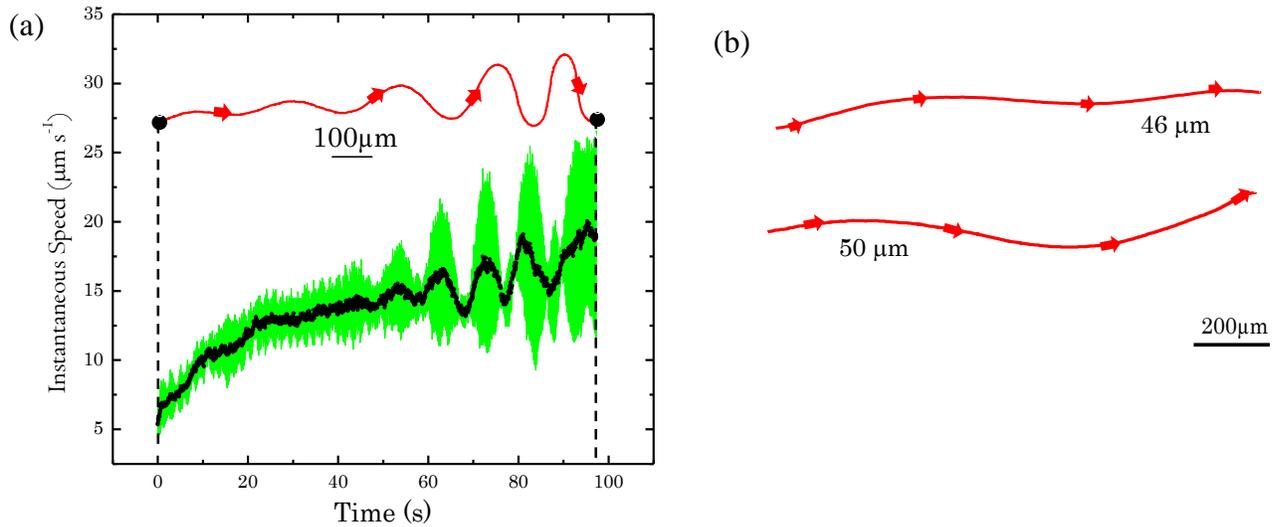

Figure 9. (a) Varation of instantaneous speed of an active droplet performing rheotaxis soon after the flow is started at $t=0$. (*channel height h*= 3.2mm) The inset shows the correspoding X-Y trajectory. (b) X-Y trajectories of smaller droplets of size 46 µm and 50 µm performing upstream rheotaxis with lesser curvature in their oscillations.

To further verify this claim, in figure 9(a), we show the time evolution of the average instantaneous speed of a representative rheotactic trajectory soon after the external flow is started ($t=0$) and the inset shows the corresponding X-Y trajectory. Clearly, the trajectory begins with a lower speed causing weaker oscillations in the initial part of the trajectory. As time progresses, the droplet accelerates resulting in enhanced oscillations. We also repeated the experiment to probe the rheotactic motion of smaller ~ (45 µm and 50 µm) droplets, which under similar shear rate (~12 ml hr$^{-1}$) demonstrated significantly lesser oscillatory behavior, see figure 9(b). Here, despite similar $|V_{rel.}|$ for the droplet, due to their smaller size, stronger elastic forces ($\sim 1/R^2$) result in smaller $\theta_s$, and therefore lesser oscillations. These observations provide further evidence towards the nematic phase being the root cause behind the oscillatory rheotaxis of active droplets.

## 4. Conclusions

In this paper, for the first time, to the best our knowledge, we demonstrate the upstream rheotactic behavior of active droplets, when exposed to external shear flow near a planar wall. The effect originates due to the flow gradient across the droplet generating an anisotropy in the concentration gradient of surfactant/filled micelles in the flow direction, which otherwise remains isotropic in quiescent ambient conditions. Our results suggest that at low shear rates, this alteration by shear flow remains weak and droplets continue their random active motion. At very high flow rates, the flow washes out the filled micelles, snuffing out the required concentration gradient for active motion, and the droplets simply drift with the flow in the downstream direction. Away from the wall as well, the external flow does not generate enough anisotropy, and the droplets either perform random active motion or drift downstream. It is only under moderate flow rates near the wall that the anisotropy in the concentration gradient is successfully established, which results in rheotaxis. Another key observation is that while performing rheotaxis, the trajectories of active droplets are oscillatory in nature, which is attributed to the nematic phase of the droplet. In the end, exploiting the rheotactic behavior, we demonstrate a unique capability to hydrodynamically trap as well as shuttle active droplets, a phenomenon with promising potential for novel biomedical applications. Deeper insights into this unique behavior in future, both through experiments and analytical studies, will facilitate enhanced capabilities of controlled drug delivery systems.

**Supplementary Material**

Calculation of theoretical fluid flow profile in the optical cell, Velocity autocorrelation decay with time for 5CB active droplets in 6wt% TTAB aqueous solution without external flow.

**Author's Contribution**

PD, DSP and RM conceptualized the research, PD & RM designed the methodology of the experiments, PD performed the experiments, PD, As and RM analyzed the experimental data, PD, DSP and RM wrote the paper.

**Acknowledgement**

We acknowledge the funding received by the Science and Engineering Research Board (SB/S2/RJN-105/2017), Department of Science and Technology, India.

**Data Availability**

The data that support the findings of this study are available from the corresponding author upon reasonable request.

**References**

1. A. Zöttl, H. Stark, Emergent behavior in active colloids. *J. Phys. Condens. Matter*. **28** (2016),

doi:10.1088/0953-8984/28/25/253001.

2. J. Elgeti, R. G. Winkler, G. Gompper, Physics of microswimmers - Single particle motion and collective behavior: A review. *Reports Prog. Phys.* **78** (2015), doi:10.1088/0034-4885/78/5/056601.

3. T. Tran-Cong, M. Ramia, A boundary-element analysis of flagellar propulsion. *J. Fluid Mech.* **184**, 533–549 (1987).

4. M. Ramia, D. L. Tullock, N. Phan-Thien, The role of hydrodynamic interaction in the locomotion of microorganisms. *Biophys. J.* **65**, 755–778 (1993).

5. J. Hill, O. Kalkanci, J. L. McMurry, H. Koser, Hydrodynamic surface interactions enable escherichia coli to seek efficient routes to swim upstream. *Phys. Rev. Lett.* **98**, 1–4 (2007).

6. I. P. Dobrovolsky, Hydrodynamic phenomena. *Annu. Rev. Fluid Mech.* **41**, 313–58 (1992).

7. Marcos, H. C. Fu, T. R. Powers, R. Stocker, Bacterial rheotaxis. *Proc. Natl. Acad. Sci. U. S. A.* **109**, 4780–4785 (2012).

8. A. Dauptain, J. Favier, A. Bottaro, Hydrodynamics of ciliary propulsion. *J. Fluids Struct.* **24**, 1156–1165 (2008).

9. E. A. Gillies, R. M. Cannon, R. B. Green, A. A. Pacey, Hydrodynamic propulsion of human sperm. *J. Fluid Mech.* **625**, 445–474 (2009).

10. J. R. Howse, R. A. L. Jones, A. J. Ryan, T. Gough, R. Vafabakhsh, R. Golestanian, Self-Motile Colloidal Particles: From Directed Propulsion to Random Walk. *Phys. Rev. Lett.* **99**, 8–11 (2007).

11. J. L. Anderson, D. C. Prieve, Diffusiophoresis: Migration of Colloidal Particles in Gradients of Solute Concentration. *Sep. Purif. Rev.* **13**, 67–103 (1984).

12. T. Banno, A. Asami, N. Ueno, H. Kitahata, Y. Koyano, K. Asakura, T. Toyota, Deformable Self-Propelled Micro-Object Comprising Underwater Oil Droplets. *Sci. Rep.* **6** (2016), doi:10.1038/srep31292.

13. T. Ban, H. Nakata, Metal-Ion-Dependent Motion of Self-Propelled Droplets Due to the Marangoni Effect. *J. Phys. Chem. B*. **119**, 7100–7105 (2015).

14. T. Banno, R. Kuroha, T. Toyota, pH-sensitive self-propelled motion of oil droplets in the presence of cationic surfactants containing hydrolyzable ester linkages. *Langmuir*. **28**, 1190–1195 (2012).

15. T. Banno, S. Miura, R. Kuroha, T. Toyota, Mode changes associated with oil droplet movement in solutions of gemini cationic surfactants. *Langmuir*. **29**, 7689–7696 (2013).

16. S. Herminghaus, C. C. Maass, C. Krüger, S. Thutupalli, L. Goehring, C. Bahr, Interfacial mechanisms in active emulsions. *Soft Matter*. **10**, 7008–7022 (2014).

17. A. Izzet, P. G. Moerman, P. Gross, J. Groenewold, A. D. Hollingsworth, J. Bibette, J. Brujic,

Tunable Persistent Random Walk in Swimming Droplets. *Phys. Rev. X*. **10**, 1–8 (2020).

18. I. Lagzi, S. Soh, P. J. Wesson, K. P. Browne, B. A. Grzybowski, Maze solving by chemotactic droplets. *J. Am. Chem. Soc.* **132**, 1198–1199 (2010).

19. C. C. Maass, C. Krüger, S. Herminghaus, C. Bahr, Swimming Droplets. *Annu. Rev. Condens. Matter Phys.* **7**, 171–193 (2016).

20. P. G. Moerman, H. W. Moyses, E. B. Van Der Wee, D. G. Grier, A. Van Blaaderen, W. K. Kegel, J. Groenewold, J. Brujic, Solute-mediated interactions between active droplets. *Phys. Rev. E*. **96**, 1–8 (2017).

21. K. Peddireddy, P. Kumar, S. Thutupalli, S. Herminghaus, C. Bahr, Solubilization of thermotropic liquid crystal compounds in aqueous surfactant solutions. *Langmuir*. **28**, 12426–12431 (2012).

22. M. Schmitt, H. Stark, Marangoni flow at droplet interfaces: Three-dimensional solution and applications. *Phys. Fluids*. **28** (2016), doi:10.1063/1.4939212.

23. Z. Izri, M. N. Van Der Linden, S. Michelin, O. Dauchot, Self-propulsion of pure water droplets by spontaneous marangoni-stress-driven motion. *Phys. Rev. Lett.* **113**, 1–5 (2014).

24. P. Dwivedi, B. R. Si, D. Pillai, R. Mangal, Solute induced jittery motion of self-propelled droplets. *Phys. Fluids*. **33** (2021), doi:10.1063/5.0038716.

25. S. Sundararajan, P. E. Lammert, A. W. Zudans, V. H. Crespi, A. Sen, Catalytic motors for transport of colloidal cargo. *Nano Lett.* **8**, 1271–1276 (2008).

26. M. Li, M. Brinkmann, I. Pagonabarraga, R. Seemann, J. B. Fleury, Spatiotemporal control of cargo delivery performed by programmable self-propelled Janus droplets. *Commun. Phys.* **1** (2018), doi:10.1038/s42005-018-0025-4.

27. A. J. T. M. Mathijssen, N. Figueroa-Morales, G. Junot, É. Clément, A. Lindner, A. Zöttl, Oscillatory surface rheotaxis of swimming E. coli bacteria. *Nat. Commun.* **10**, 7–9 (2019).

28. S. Campuzano, J. Orozco, D. Kagan, M. Guix, W. Gao, S. Sattayasamitsathit, J. C. Claussen, A. Merkoçi, J. Wang, Bacterial isolation by lectin-modified microengines. *Nano Lett.* **12**, 396–401 (2012).

29. W. Gao, S. Sattayasamitsathit, A. Merkoc, A. Escarpa, J. Wang, Superhydrophobic Alkanethiol- Coated Microsubmarines for E ff ective Removal of Oil, 4445–4451 (2012).

30. J. Adler, Chemotaxis in Bacteria. *Science (80-. )*. **153**, 708–716 (2008).

31. B. M. Friedrich, F. Jülicher, Chemotaxis of sperm cells. *Proc. Natl. Acad. Sci. U. S. A.* **104**, 13256–13261 (2007).

32. C. Jin, C. Krüger, C. C. Maass, Chemotaxis and autochemotaxis of self-propelling droplet swimmers. *Proc. Natl. Acad. Sci.* **114**, 5089–5094 (2017).

33. T. Goto, K. Nakata, K. Baba, M. Nishimura, Y. Magariyama, A fluid-dynamic interpretation of the asymmetric motion of singly flagellated bacteria swimming close to a boundary. *Biophys. J.* **89**, 3771–3779 (2005).

34. E. Lauga, W. R. DiLuzio, G. M. Whitesides, H. A. Stone, Swimming in circles: Motion of bacteria near solid boundaries. *Biophys. J.* **90**, 400–412 (2006).

35. R. M. Macnab, Bacterial flagella rotating in bundles: A study in helical geometry. *Proc. Natl. Acad. Sci. U. S. A.* **74**, 221–225 (1977).

36. P. D. Frymier, R. M. Ford, H. C. Berg, P. T. Cummins, Three-dimensional tracking of motile bacteria on planar surface. *Proc. Natl. Acad. Sci. U. S. A.* **92**, 6195–6199 (1995).

37. F. Rühle, J. Blaschke, J. T. Kuhr, H. Stark, Gravity-induced dynamics of a squirmer microswimmer in wall proximity. *New J. Phys.* **20** (2018), , doi:10.1088/1367-2630/aa9ed3.

38. D. J. Smith, E. A. Gaffney, J. R. Blake, J. C. Kirkman-Brown, Human sperm accumulation near surfaces: A simulation study. *J. Fluid Mech.* **621**, 289–320 (2009).

39. L. J. Fauci, A. McDonald, Sperm motility in the presence of boundaries. *Bull. Math. Biol.* **57** (1995), pp. 679–699.

40. D. Giacché, T. Ishikawa, T. Yamaguchi, Hydrodynamic entrapment of bacteria swimming near a solid surface. *Phys. Rev. E - Stat. Nonlinear, Soft Matter Phys.* **82**, 1–8 (2010).

41. H. Shum, E. A. Gaffney, D. J. Smith, Modelling bacterial behaviour close to a no-slip plane boundary: The influence of bacterial geometry. *Proc. R. Soc. A Math. Phys. Eng. Sci.* **466**, 1725–1748 (2010).

42. Z. Zhang, J. Liu, J. Meriano, C. Ru, S. Xie, J. Luo, Y. Sun, Human sperm rheotaxis: A passive physical process. *Sci. Rep.* **6**, 1–8 (2016).

43. G. Jing, A. Zöttl, É. Clément, A. Lindner, Chirality-induced bacterial rheotaxis in bulk shear flows. *Sci. Adv.* **6** (2020), doi:10.1126/sciadv.abb2012.

44. F. P. Bretherton, Rheotaxis of spermatozoa. *Proc. R. Soc. London. Ser. B. Biol. Sci.* **153**, 490–502 (1961).

45. R. Rosengarten, A. Klein-Struckmeier, H. Kirchhoff, Rheotactic behavior of a gliding mycoplasma. *J. Bacteriol.* **170** (1988), pp. 989–990.

46. J. Palacci, S. Sacanna, A. Abramian, J. Barral, K. Hanson, A. Y. Grosberg, D. J. Pine, P. M. Chaikin, Artificial rheotaxis. *Sci. Adv.* **1**, 1–7 (2015).

47. R. Baker, J. E. Kauffman, A. Laskar, O. E. Shklyaev, M. Potomkin, L. Dominguez-Rubio, H. Shum, Y. Cruz-Rivera, I. S. Aranson, A. C. Balazs, A. Sen, Fight the flow: The role of shear in artificial rheotaxis for individual and collective motion. *Nanoscale*. **11**, 10944–10951 (2019).

48. L. Ren, D. Zhou, Z. Mao, P. Xu, T. J. Huang, T. E. Mallouk, Rheotaxis of Bimetallic Micromotors Driven by Chemical-Acoustic Hybrid Power. *ACS Nano*. **11**, 10591–10598 (2017).
49. J. Katuri, W. E. Uspal, J. Simmchen, A. Miguel-López, S. Sánchez, Cross-stream migration of active particles. *Sci. Adv.* **4**, 1–12 (2018).
50. B. R. Si, P. Patel, R. Mangal, Self-Propelled Janus Colloids in Shear Flow. *Langmuir*. **36**, 11888–11898 (2020).
51. T. Toyota, N. Maru, M. M. Hanczyc, T. Ikegami, T. Sugawara, Self-propelled oil droplets consuming "Fuel" surfactant. *J. Am. Chem. Soc.* **131**, 5012–5013 (2009).
52. C. Jin, C. Kru-ger, C. C. Maass, Chemotaxis and autochemotaxis of self-propelling droplet swimmers. *Proc. Natl. Acad. Sci. U. S. A.* **114**, 5089–5094 (2017).
53. M. M. Hanczyc, T. Toyota, T. Ikegami, N. Packard, T. Sugawara, Fatty acid chemistry at the oil-water interface: Self-propelled oil droplets. *J. Am. Chem. Soc.* **129**, 9386–9391 (2007).
54. M. Suga, S. Suda, M. Ichikawa, Y. Kimura, Self-propelled motion switching in nematic liquid crystal droplets in aqueous surfactant solutions. *Phys. Rev. E*. **97**, 1–8 (2018).
55. V. Kantsler, J. Dunkel, M. Blayney, R. E. Goldstein, Correction: Rheotaxis facilitates upstream navigation of mammalian sperm cells. *Elife*. **3**, e03521 (2014).
56. S. Das, S. Manda, S. Chakraborty, Cross-stream migration of a surfactant-laden deformable droplet in a Poiseuille flow. *Phys. Fluids*. **29** (2017), doi:10.1063/1.4997786.
57. O. S. Pak, J. Feng, H. A. Stone, Viscous Marangoni migration of a dropl in a Poiseuille flow at low surface Peclet numbers. *J. Fluid Mech.* **753**, 535–552 (2014).
58. C. Krüger, G. Klös, C. Bahr, C. C. Maass, Curling Liquid Crystal Microswimmers: A Cascade of Spontaneous Symmetry Breaking. *Phys. Rev. Lett.* **117**, 1–5 (2016).
59. M. Joly, Nature 27. **228**, 4001 (1946).